Densest Subgraphs

# THE DENSE K SUBGRAPH PROBLEM

Author: Goldstein Doron

ID: 028672129

Supervisor: Dr. Michael Langberg

Jan, 2010

Israel

# THE DENSE K SUBGRAPH PROBLEM

ABSTRACT. Given a graph $G = (V, E)$ and a parameter $k$, we consider the problem of finding a subset $U \subseteq V$ of size $k$ that maximizes the number of induced edges (D$k$S). We improve upon the previously best known approximation ratio for D$k$S, a ratio that has not seen any progress during the last decade. Specifically, we improve the approximation ratio from $n^{0.32258}$ to $n^{0.3159}$. The improved ratio is obtained by studying a variant to the D$k$S problem in which one considers the problem of finding a subset $U \subseteq V$ of size *at most $k$* that maximizes the number of induced edges. Finally, we study the D$k$S variant in which one considers the problem of finding a subset $U \subseteq V$ of size *at least $k$* that maximizes the number of induced edges.

## CONTENTS

## 1. Introduction

In this thesis, we consider the Densest $k$ Subgraph problem (D$k$S). For a given undirected graph instance $G(E, V)$, in the D$k$S problem one is to find a subgraph with exactly $k$ vertices with a maximum number of induced edges. In addition, we also consider two variants of D$k$S. The Densest-at-least-$k$-Subgraph problem and the Densest-at-most-$k$-Subgraph problem (both defined below). We present a number of results for the problems at hand. Most notably, we improve upon the previously best known approximation ratio for D$k$S, a ratio that has not seen any progress during the last decade.

### 1.1. **Definitions.**

**Definition 1.1** (Densest-$k$-Subgraph). Given an undirected graph $G(V, E)$ the Densest-$k$-Subgraph (D$k$S) problem on $G$ is the problem of finding a subset $U \subseteq V$ of vertices of size $k$ with the maximum induced average degree. The average degree of the optimal subgraph will be denoted as $d^* = 2|E(U)|/k$. Here $|E(U)|$ denotes the number of edges in the subgraph induced by $U$.

**Definition 1.2** (Densest-at-least-$k$-Subgraph). Given an undirected graph $G(V, E)$ the Densest-at-least-$k$-Subgraph (Dal$k$S) problem on $G$ is the problem of finding a subset $U \subseteq V$ of vertices of size *at least* $k$ with the maximum induced average degree as defined in the D$k$S problem.

**Definition 1.3** (Densest-at-most-$k$-Subgraph). Given an undirected graph $G(V, E)$ the Densest-at-most-$k$-Subgraph (Dam$k$S) problem on $G$ is the problem of finding a subset $U \subseteq V$ of vertices of size *at most* $k$ with the maximum induced average degree (as defined in the D$k$S problem).

An $\alpha \geq 1$ approximation algorithm for these problems is an algorithm that given $G$ returns a subset of vertices $U$ of size $k^*$ such that $2|E(U)|/k^* \geq d^*/\alpha$ where $k^* = k$ for D$k$S, $k^* \geq k$ for Dal$k$S and $k^* \leq k$ for Dam$k$S. Here, $d^*$ is the average degree of the optimal subgraph on each of the problems respectively.

### 1.2. **Previous work.**
The D$k$S problem is NP-hard to solve exactly (a fact easily seen by a reduction from the Max-Clique problem). The current best approximation ratio known for the D$k$S problem is $n^\delta$ for some $\delta < \frac{1}{3}$ [5], where $n$ is the number of vertices in the input graph. This result was obtained using a combinatorial algorithm. To be more precise, the algorithm of [5] is actually combined from five different combinatorial algorithms, each of the algorithms gives good results on different instances of the problem. The first algorithm is a trivial one which always returns a subgraph with average degree of value at least 1. The second algorithm is greedy and performs well when the input graph has several vertices with high average degree relative to $n$. The third algorithm is also greedy, and gives good results when $d^*$, the optimum value of the D$k$S instance, is high with respect to its size $k$. The final two algorithms are tailor made to fit specific relations between $d^*$ and $k$.

In [5] it is shown that it suffices to consider the first three algorithms to obtain a ratio of $n^{\frac{1}{3}}$. Improving the ratio to $n^{1/3-\epsilon}$, for some constant $\epsilon > 0$, is obtained by combining the two additional algorithms that are designed to take care of the special instances for which the first three algorithms indeed achieve a ratio no better than $n^{1/3}$. The exact value of $\epsilon$ achieved in [5] it not stated explicitly, rather it is only shown to be constant and thus independent of $n$. Nevertheless, since the work of [5] (over a decade ago), there has been no improvement in the approximation ratio of the D$k$S problem. In this thesis, we present a detailed analysis of the ratio obtained by [5] (Section 5), and improve on this ratio by replacing the fifth algorithm of [5] with a new one (Section 6).

The D$k$S problem was also studied by Feige and Seltser [7] where it is shown to be NP-complete even when restricted to bipartite graphs of maximum degree 3 (use a reduction from the Max-Clique problem). In a similar way Asahiro, Hassin, and Iwama [2] have showed that the problem remains NP-complete in very sparse graphs where $d = k^\epsilon$.

Khot [9] has shown there can be no PTAS solution for the densest k-subgraph problem, under the assumption that the family NP does not have randomized algorithms that run in time $2^{n^\epsilon}$ for some constant $\epsilon > 0$.

Two additional problems studied in this thesis, that are closely related to D$k$S and first appear in the work of Anderson [1], are the Dal$k$S and Dam$k$S problems (Definitions 1.2 and 1.3 respectively). For the Dal$k$S problem, Anderson presents an approximation algorithm with a ratio of 2. The question whether Dal$k$S is NP-hard or not was not resolved in [1]. This is not the case for Dam$k$S, as in [1], Anderson proves that Dam$k$S is NP hard (by a simple reduction to the Max-Clique problem). Moreover, he presents a connection between approximating Dam$k$S and D$k$S. Namely, if there exists a polynomial time algorithm that approximates Dam$k$S in a weak sense, returning a set of at most $\beta k$ vertices with average degree at least $1/\gamma$ times the average degree of the densest subgraph on at most $k$ vertices, then there exists a polynomial time approximation algorithm for D$k$S with ratio $4(\gamma^2 + \gamma\beta)$.

Another problem closely related to the D$k$S problem is the Densest Subgraph (DS) problem. The DS Problem concerns choosing a subset $U$ (regardless of its size) as to maximize the average degree of the subgraph induced bu $U$. The DS problem can be solved in polynomial time using either LP based techniques [4] or by flow techniques [8].

We note that a recent and independent work [10] presents results similar to ours regarding the Dal$k$S and the Dam$k$S problems. Moreover, in the recent and independent work [3] a better overall approximation guarantee of $n^{1/4+\epsilon}$ is given for the D$k$S problem. The running time in [3] depends on $\epsilon$, and to obtain a ratio of $n^{1/4}$ a total running time of $n^{O(logn)}$ is needed.

### 1.3. Our results.

In this thesis, we present several results regarding the D$k$S and the closely related problems of Dam$k$S and Dal$k$S.

Our main result is the design and analysis of a new algorithm to be used in combination with the first four algorithms of [5] (replacing the fifth algorithm of [5]). We show that adding our algorithm one is able to obtain an approximation ratio of $n^{0.3159}$ for D$k$S. To compare this with the ratio of [5], we first study the original algorithms of [5] and show that their combination results in an approximation ratio of $n^{0.32258} = n^{1/3-\epsilon}$ (thus computing the value of $\epsilon$ unspecified in [5]). Our new algorithm is based on linear programming (LP), and is designed to improve on the quality of the algorithms of [5] on certain D$k$S instances. The exact ratio (or to be more precise an upper bound on the ratio) of the combined five algorithms is obtained using a simple C-program.

The principle of our new algorithm is to guess $d^*$ (the optimum average degree), guess a vertex $v$ inside an optimal solution (an optimal subset $U$), and then run an LP that takes $d^*$ and $v$ into account. In our LP, a suitable D$k$S solution corresponds to a feasible LP solution, but the other direction does not necessarily hold. Thus, the fractional LP solution is rounded to get a feasible solution $U$ to D$k$S. The resulting approximation ratio obtained depends on several parameters of the instance at hand. Specifically, the ratio depends on $k$, $d^*$, and $d_H$ the maximum degree in $G$ and is equal to $r_6 = \left( \frac{d_H^4 k}{(d^*)^4} \right)^{\frac{1}{5}}$. As mentioned, integrating our new algorithm with the four of [5] (replacing the fifth one with ours) we obtain the improved ratio.

Our other results were motivated by the work of Anderson [1]. In [1] the Dal$k$S problem was presented (see Definition 1.2), but the NP-hardness of this problem was left open. We have managed to show, using a reduction from the D$k$S problem, that Dal$k$S is also NP-hard. Our proof is based on taking a D$k$S instance and adding to it a big clique, making it an instance for the Dal$k$S problem. Also in [1], a 2-approximation was given to solve the Dal$k$S problem. We have used [11] to show a more general approach for solving this problem based on optimizing supermodular functions. In [1] it was shown that if Dam$k$S (see Definition 1.3) has a $\gamma$-approximation agorithm, then the algorithm can be used to approximate D$k$S within a ratio of $4\gamma^2$. We reduce the latter to $4\gamma$. Our improved reduction is iterative and is strongly based on the fact that in each iteration we only remove edges from the original graph $G$ if they are included in the final D$k$S solution.

1.4. **Thesis structure.** Our work includes the following sections. First we prove that Dal$k$S is NP-complete (Section 2). This resolves the question left open in [1]. We proceed to present a general paradigm that enables to obtain a 2 approximation algorithm for Dal$k$S in polynomial time (Section 3). This complements the 2-approximation algorithm presented in [1]. Turning to the Dam$k$S problem, in Section 4 we show a strong connection between the ability to approximate Dam$k$S and D$k$S. Namely, improving on the connections discribed in [1] we show that any ratio achievable on Dam$k$S is also (up to constant factors) achievable on D$k$S. In Section 5, we study the D$k$S problem, and present a full analysis of the approximation ratio of [5]. Namely, we compute the value of $\epsilon$ left unresolved in [5]. In Section 6 we present our new algorithm for D$k$S. This section strongly builds on the previous ones. Finally, in Section 7 we present the numerical techniques (a C-program) used to determine the ratios stated throughout our work. We also rigorously bound any slackness that may arise from these techniques.

## 2. THE DAL$k$S PROBLEM IS NP HARD

In his work [1], Anderson gave a 2-approximation algorithms for the Dal$k$S problem and mentioned that he doesn't know if this problem is NP hard or not. We have found that this problem is NP hard.

**Theorem 2.1.** *Finding the densest subgraph with at least $k$ vertices is NP hard.*

*Proof.* In this proof, for a graph $H$, the term *density* will refer to the average degree of $H$. Let $G(V, E)$ be an instance to the D$k$S problem. Let $|V| = n$. Let $G'$ be the graph consisting of $G$ and a clique of size $3n$: $K_{3n}$. Namely, the vertex set of $G'$ consists of the vertices of $G$ and an additional $3n$ new vertices; and the edge set of $G'$ consists of the edges of the original graph $G$ and the edges of a complete graph on the new $3n$ vertices. Let $k' = k + 3n$. Let $H$ be the optimal subset in $G'$ with respect to the Dal$k$S problem with parameter $k'$. We claim that $H$ will consist of exactly the densest subgraph in $G$ of size $k$ and the clique $K_{3n}$. This will suffice to prove our claim (as the D$k$S problem is NP-hard).

First we show that $K_{3n} \subseteq H$. Suppose that the size of $H$ is $b_1 + b_2$ where $b_1$ denotes the number of vertices in $H$ taken from the clique $K_{3n}$ and $b_2$ denotes the number of vertices in $H$ taken from the original graph $G$. For the sake of contradiction assume $b_1 < 3n$. We will show that we can take $3n - b_1$ vertices out of the $b_2$ vertices in $G$ and select instead $3n - b_1$ vertices in $K_{3n}$ making the clique complete. By doing so we will increase the number of edges in $H$ without increasing the number of vertices, hence increasing the density and contradicting the optimality of the solution.

So lets show that the number of edges increases. First we give names to the different groups of vertices. Group $B1$ will be the $b_1$ vertices of $K_{3n}$ defined above, $B2$ will be the $b_2$ vertices of $G$ defined above, $T1$ will be the group of new selected vertices in $K_{3n}$ (namely $T1 = K_{3n} \setminus B1$) , $T2$ will be the $3n - b_1$ vertices we removed from $B2$, and $R2 = B2 \setminus T2$. We need to show that $|E(T1)| + |\delta(T1, B1)| > |E(T2)| + |\delta(T2, R2)|$. Here $\delta(A, B)$ refers to the edges crossing between vertex sets $A$ and $B$. Notice that $b_1 + b_2 > 3n$.

Since $T1$ is a subset of the clique $K_{3n}$ and $|T1| = |T2|$ it holds that $|E(T1)| \geq |E(T2)|$. Next we want to show that $|\delta(T1, B1)| > |\delta(T2, R2)|$. $T1$ and $B1$ are subsets of the clique $K_{3n}$, thus there are edges between all the vertices in $T1$ and all the vertices in $B1$. Since $T1$ has the same size as $T2$, the only way $|\delta(T2, R2)|$ will be greater or equal to $|\delta(T1, B1)|$ is if $|R2|$ is greater or equal to $|B1|$. We now show that this cannot happen, namely $|B1| > |R2|$. $|B1| = 3n - |T1| > n - |T2| \geq |B2| - |T2| = |R2|$. Thus, we deduce that every optimal solution $H$ must include the clique $K_{3n}$.

Next we will show that exactly $k$ vertices will be selected from $G$. Recall that the number of vertices chosen from $G$ is denoted $b_2$. $b_2$ must be at least $k = k' - 3n$. This follows from the fact that the algorithm returns a subgraph $H$ of $G'$ of size at least $k'$ which includes the clique $K_{3n}$ and additional vertices from $G$. Now we show that selecting more then $k'$ vertices from $G$ will contradict the optimality of the solution.

Suppose for the sake of contradiction that the solution $H$ has $b_2$ vertices from $G$ where $b_2 > k$. Let $d$ be the average degree of the subgraph of $G$ induced by these $b_2$ vertices. The average degree of the resulting graph $\bar{d}$ will be $\bar{d} = \frac{3n(3n-1)+db_2}{3n+b_2}$. If we remove the vertex of minimal degree in $H \cap G$ we will loose at most $d$ edges (or, at most $2d$ in the sum of degrees), hence getting a new average degree $\hat{d} \geq \frac{3n(3n-1)+d(b_2-2)}{3n+b_2-1}$. This new density is bigger than $\bar{d}$ as can be seen from the following calculation. $\hat{d} - \bar{d} = \frac{3n(3n-1)+d(b_2-2)}{3n+b_2-1} - \frac{3n(3n-1)+db_2}{3n+b_2} = \frac{9n^2-3n-6nd-b_2d}{(3n+b_2)(3n-1+b_2)} > 0$. As, $d \leq n$ and $b_2 \leq n$, it holds that $\hat{d} - \bar{d} \geq \frac{2n^2-3n}{(3n+b_2)(3n-1+b_2)} > 0$. The last inequality stands for $n > 1$. So we found a way to improve the density by removing vertices from $H$. This implies that $b_2$ must be as small as possible. Nevertheless the demand is that it will be at least $k'$, so we deduce $b_2 = k$.

Finally, the $k$ vertices from $G$ that will be selected for $H$ will contribute the most when they consist of the densest subgraph of $G$ with size $k$. Let $d$ be the average degree of the $k$ vertices selected from $G$. Let $d^* > d$ be the average degree of the densest subgraph of size $k$ in $G$. As $\bar{d} = \frac{3n(3n-1)+kd}{k'} < \frac{3n(3n-1)+kd^*}{k'}$. We gain from taking the densest subgraph of size $k$ in $G$.

We conclude that the $k$ vertices selected from $G$ must be the D$k$S solution. Since D$k$S is NP hard we conclude our assertion. □

## 3. A 2 approximation algorithm for the Dal$k$S problem

In his work [1], Anderson presented a 2 approximation for the Dal$k$S problem. We present an alternative proof that holds for a family of problems which includes Dal$k$S. Our algorithm is based on the algorithm for the Dense-Subgraph problem presented in [11].

Let $G(V, E)$ be an undirected graph and let $S$ be any subset of $V$. Denote by $E(S)$ the edges induced by the subset $S$.

**Lemma 3.1.** *Let $q > 0$, we can solve the problem of maximizing $|E(S)| - q|S|$ exactly in polynomial time.*

*Proof.* Let $U$ be a ground set, and let $X$ and $Y$ be subsets of $U$. A set function $f$ is supermodular if

$$f(X) + f(Y) \leq f(X \cap Y) + f(X \cup Y) \quad \forall X, Y \subseteq U$$

A set function $p$ is submodular if

$$p(X) + p(Y) \geq p(X \cap Y) + p(X \cup Y) \quad \forall X, Y \subseteq U$$

If $f(X)$ is supermodular and $p(X)$ is submodular, then for $q > 0$ the function $f(X) - qp(X)$ is supermodular. The problem of maximizing a supermodular function can be solved (under certain *rationality* restrictions) in polynomial time (e.g., see [11]).

For $G(V, E)$, let $X$ and $Y$ be subsets of $V$. We now show that the function of $\gamma(X) = |E(X)|$ is supermodular, and the function $\delta(X) = |X|$ is submodular. This follows by counting the contribution of edges to the sides in the definition of $f$ above. Edges in $E(X - Y)$ and $E(Y - X)$ are counted once in both sides, while edges in $E(X \cap Y)$ are counted twice. Edges between $X - Y$ and $Y - X$ are counted only on the right hand side. This proves our claim for $\gamma(X)$. The proof for $\delta(X)$ is straightforward. □

Maximizing $|E(S)| - q|S|$ for any $q > 0$ can aid us in finding a 2 approximation for Dal$k$S. Let us define $G^*(S^*, E^*)$ as the densest subgraph with at least $k$ vertices, and its average degree is $d^*$.

Namely, $|E^*| = d^*|S^*|/2$. For $q = d^*/4$, let $S$ be the subset maximizing $|E(S)| - q|S|$ and let $|E(S)| = d|S|/2$. We show that $S$ will imply our approximate solution. It holds that:

$$|E(S)| - \frac{d^*}{4}|S| \geq |E(S^*)| - \frac{d^*}{4}|S^*| = |E^*| - \frac{d^*}{4}|S^*| = |E^*|/2$$

If $|S| \geq |S^*|$ from the fact that $|E(S)| - \frac{d^*}{4}|S| \geq 0$ we get $d|S|/2 = |E(S)| \geq \frac{d^*}{4}|S|$ which implies that $d \geq d^*/2$.

If $|S| < |S^*|$ we add arbitrary vertices to $S$ until it is of size $k$. Let $E'$ be the edge set induced by the enlarged set $S$ and let $d' = 2|E'|/k$ be its average degree.

We have:

$$d' = 2|E'|/k \geq 2|E(S)|/k \geq |E^*|/k \geq |E^*|/|S^*| \geq d^*/2$$

The third inequality follows from the fact that $|E(S)| - \frac{d^*}{4}|S| \geq |E^*|/2$. We thus achieve a 2 approximation no matter what the size of $S$ is.

*Remark* 3.2. Since we don't know $d^*$, we can't compute $q$ in advance. We can try to guess it in different ways. The naive approach will be to exhaust all the possibilities. Since $d^* = 2|E^*|/|S^*|$ and $|E^*| \in \{0, 1, ..., |E|\}$ and $|S^*| \in \{0, 1, ..., |S|\}$ then we can bound the number of possibilities for $d^*$ by with $|E||S|$.

## 4. The Dam$k$S and D$k$S problems

**Definition 4.1.** An algorithm $A(G, k)$ is a $(\beta, \gamma)$-approximation algorithm for the Dam$k$S problem if for input graph $G$ and size $k$ it returns a solution with at most $\beta k$ vertices and an average degree of at least $dam(G, k)/\gamma$, where $dam(G, k)$ is the optimal average degree of the Dam$k$S problem on $G$.

Anderson, in his work [1], has shown the following: If there is a $(\beta, \gamma)$-approximation algorithm for Dam$k$S (with $\beta$ and $\gamma$ greater or equal than 1), then there is a $4(\gamma^2 + \gamma\beta)$-approximation algorithm for D$k$S. For the specific case where $\beta = 1$ this gives a $4\gamma^2$-approximation ratio. We significantly improve upon this result of [1]:

**Theorem 4.2.** *If there is a $\gamma$-approximation algorithm for DamkS (with $\gamma$ greater or equal than 1) then there is a $4\gamma$-approximation algorithm for DkS.*

*Proof.* We specify our algorithm for D$k$S. As done in several places before in this thesis, we assume that the value $d^*$ is known.

**Algorithm 4.3.** [Solve D$k$S using Dam$k$S]

*Let $G(V, E)$ and $k$ be an input instance to the DkS problem. Let $S$ be an empty group of vertices.*

*a) Using the approximation algorithm for DamkS on $G$, find an approximate solution $S'$ with at most $k$ vertices.*

*b) Add the vertices of $S'$ and its induced edges to $S$. Remove the edges induced by $S'$ from $G$.*

*c) If the number of edges in $S$ is $E(S) < \frac{1}{4}kd^*$ and $|S| < k$ we go back to (a) and continue.*

*d) If $|S| < k$ we add to it an arbitrary set of $k - |S|$ additional vertices out of $G$.*

*e) Otherwise for $|S| > k$. We greedily remove the lowest degree vertices from $S$ until it is of size $k$.*

*f) Return $S$.*

We now analyze the suggested algorithm:

Denote an optimal subset to the D$k$S problem of size $k$ by $S^*$, and its edge set by $E^*$. It holds that $d^*k = 2|E^*|$ for the optimal average degree $d^*$. If at the end of our algorithm $S$ includes at least $\frac{1}{4\gamma}|E^*|$ edges of $E$ then it holds that $|E(S)| \geq \frac{1}{8\gamma}kd^*$. As $|S| = k$ this implies an approximation ratio of $4\gamma$.

We will prove that each iteration of (a) picks vertices with average degree of value at least $\frac{1}{2\gamma}d^*$. In the first time we are at (a) the graph $G$ is still the original graph and so is the optimal subset $S^*$, so the Dam$k$S algorithm can pick a subgraph smaller or equal to $k$ with average degree equal or higher then $d^*/\gamma$.

**Lemma 4.4.** *At any other iteration, one of the two must exist: either the vertices of $S^*$ in $G$ still have average degree higher than $\frac{1}{2}d^*$ or the set $S$ satisfies $|E(S)| \geq \frac{1}{4}kd^*$.*

*Proof.* Suppose the second condition does not hold, then $|E(S)| < \frac{1}{4}kd^*$. This means that $S$ includes less then half of the edges that were originally induced by $S^*$. It follows that there are yet more then $\frac{1}{4}kd^*$ edges between the vertices of $S^*$. So the set $S^*$ in $G$ has at most $k$ vertices while having at least $\frac{1}{4}kd^*$ edges, namely its average degree is higher then $\frac{1}{2}d^*$. $\qquad\square$

Notice that Lemma 4.4 implies that each time we do not pass from step (c) of the algorithm to step (d) (and rater return to (a)), the set $S'$ will have average degree at least $\frac{1}{2\gamma}d^*$. This follows from the fact that there is still a subset (the subset $S^*$) in $G$ that has size at most $k$ and average degree at least $\frac{1}{2}d^*$. This implies that at each visit in step (c) the set $S$ has average degree at least $\frac{1}{2\gamma}d^*$.

By the time we pass step (c) one of the two following cases must hold. Either, $|S| < k$ and thus it must hold that their are at least $\frac{1}{4}kd^*$ edges in $S$. In this case phase (d) of the algorithm suffices to obtain the desired set $S$. Or we are in the case in which $|S| \geq k$. In this case, it holds that $|S| \leq 2k - 1$ (as in the previous iteration $S$ was smaller than $k$, and in each iteration at most $k$ vertices are added to $S$). Moreover, as in each iteration the average degree of $S'$ added to $S$ is at least $\frac{1}{2\gamma}d^*$, the average degree in $S$ is also at least $\frac{1}{2\gamma}d^*$. Thus by Lemma 4.5 (appearing below), after we remove vertices from $S$ in step (e) we remain with a set $S$ of average degree at least $\frac{1}{4\gamma}d^*$.

The following lemma that is needed for the completeness of this proof is taken from [6] (Lemma C.1). We have added it here without its proof.

**Lemma 4.5.** *(Fixing Lemma) Given a set $U$ of size $|U| > k$ and weight $W$ we can efficiently find a subset $U_k \subseteq U$ of size $k$ and weight at least $W \frac{k(k-1)}{|U|(|U|-1)}$.*

In the above, the term weight refers to the weight given to each edge in the graph and $W$ refers to the sum of these weights. In an unweighted graph, every edge has a weight of 1 and $W$ translates to the number of edges.

*Remark* 4.6. Theorem 4.2 states a connection between the Dam$k$S and D$k$S problems. Namely, a $\gamma$ approximation algorithm for the former yields a $4\gamma$ approximation for the latter. In the remainder of this thesis we will use Theorem 4.2 a few times when $\gamma = \gamma(d^*_{am})$ is a monotone decreasing function of $d^*_{am}$ - the average degree of the optimal solution to the Dam$k$S problem. For an optimal value $d^*$ to the D$k$S problem, Lemma 4.4 above implies that during any execution of step (a) in Algorithm 4.3, any subgraph $G$ considered will have a subgraph of size at most $k$ with average degree at least $d^*/2$. Note that this implies that the optimal solution to Dam$k$S on $G$ will also have average degree at least $d^*/2$. This implies, following the analysis above, that we can promise an approximation ratio for D$k$S of at most $4\gamma(d^*/2)$. The dependence of $\gamma$ on $d^*$ that we will use in the upcoming sections is polynomial, thus in these cases we obtain a slightly weaker reduction between Dam$k$S and D$k$S, however we stress that the ratio between the approximation of the former and latter remains constant in these cases.

## 5. The previously best known ratio for D$k$S

As mentioned previously the currently known best approximation ratio for the Dense-$k$-Subgraph problem stands on $n^\delta$ for a constant $\delta$ slightly less than $1/3$, [5]. The algorithm presented in [5] is composed of 5 algorithms $A_1, \ldots, A_5$. Computing the approximation ratio for each of these algorithms, [5] are able to prove that $\delta \le 1/3 - \epsilon$ for some $\epsilon > 0$. However, the analysis in [5] does not make an attempt to find the precise value of $\delta$. In what follows we calculate $\delta$. The bound, is based on the analysis of [5] and is done in two steps. First, we revisit the 5 algorithms appearing in [5] and (by refining the analysis of [5]) we present their approximation ratio as a function of $k$, $d^*$, and $d_H$. Here, as in [5], $d^*$ is the average degree of the optimal solution to the problem at hand, and $d_H$ is the average degree of the highest $k/2$ degrees in the graph. A full analysis as described above appears in [5] for the first three algorithms, so here we just state their results. For algorithms $A_4$ and $A_5$ our analysis is new. Secondly, after we have determined the approximation ratio for all five algorithms, denoted $r_i(k, d^*, d_H)$ for algorithm $A_i$, we run a simple C-program that computes (a lower bound) to

$$\max_{k, d^*, d_H} \min_{i=1,\ldots,5} r_i(k, d^*, d_H).$$

Specifically, our C-program (given in Section 7.1) computes $\min_{i=1,\ldots,5} r_i(k, d^*, d_H)$ on a large set of triplets $(k, d^*, d_H)$. Taking the maximum value of $\min_{i=1,\ldots,5} r_i(k, d^*, d_H)$ over the triplets considered yields the desired lower bound on $\delta$.

Before we state and prove the main theorem and lemmas of this section, a few remarks are in place. In [5], five algorithms were considered, and for each such algorithm $A_i$ an approximation ratio $r_i$ was determined. As stated above, we follow the proof of [5] and present an enhanced analysis for $r_i$. Using this analysis, we compute the approximation ratio obtained via combining these five algorithms. Given that our understanding of [5] is precise (we have made every effort to justify this assumption), our analysis highlights the limitation of the proof in [5] and yields Theorem 5.1 stated below. Namely, in our analysis, we present triplets $(k, d^*, d_H)$ for which the enhanced analysis of [5] does not promise an approximation ratio better than $n^{0.32258}$. We stress that our claim is on the analysis of [5] alone and not on the actual performance of their combined algorithm - which potentially could do better than the analysis shows. Throughout this section, when we state that an approximation ratio $r_i$ is equal to some expression, we mean that based on our enhanced analysis a ratio of $r_i$ can be obtained, and our analysis does not promise any ratio better than $r_i$.

**Theorem 5.1.** *Our enhanced analysis to the algorithms presented in [5] promise an approximation ratio no lower than $n^{0.32258}$.*

### 5.1. **Algorithms $A_1$, $A_2$, and $A_3$.**

**Lemma 5.2** ([5])**.** *The approximation ratio of Algorithm $A_1$ from [5] is $r_1(k, d^*, d_H) = d^*$.*

**Lemma 5.3** ([5])**.** *The approximation ratio of Algorithm $A_2$ from [5] is $r_2(k, d^*, d_H) = \frac{2nd^*}{kd_H}$.*

In [5], algorithm $A_2$ was used in more then one way. The following lemma (essentially proven in [5]) makes use of $A_2$ and is needed later.

**Lemma 5.4.** *Let $G(V, E)$ be a given graph. Let $d_H$ be the average degree of the $k/2$ highest degree vertices in $G$. One can efficiently find a subgraph $G'$ of $G$ of maximum degree $d_H$ such that either (a) Running algorithm $A_2$ on $G$ yields an approximation ratio of 3, or (b) The optimal value of the DkS problem on $G'$ is at least a third of the optimal value on $G$.*

*Proof.* Based on Lemma 3.3 in [5], removing the $k/2$ vertices with highest degree from $G$ reduces the optimum solution $d^*$ to $d' \geq d^* - 2d^*/r_2$ where $d^*/r_2$ is the average degree yielded by the second algorithm of [5]. So either $d^*/r_2 \geq d^*/3$ or $d' \geq d^*/3$ and the lemma follows. $\qquad\square$

Since a constant lost of proximity is not of our concern in this thesis, we will refer from this point on to $d_H$ as the highest degree in the graph.

**Lemma 5.5** ([5])**.** *The approximation ratio of Algorithm $A_3$ from [5] is $r_3(k, d^*, d_H) = \frac{2max(k, d_H)}{d^*}$.*

### 5.2. **Algorithm $A_4$.**

**Lemma 5.6** ([5])**.** *The approximation ratio of Algorithm $A_4$ from [5] is $r_4(k, d^*, d_H) = \frac{2k^2(2d_H)^{1/3}}{(d^*)^2}$.*

Algorithm $A_4$ in [5] activates algorithms $A_1$, $A_2$, and $A_3$ on a subgraph $G'$ of $G$ of size $n' \leq 2d_H$. By the analysis in [5] (Lemma 4.4 of [5]), $G'$ is promised to have a subgraph of size $k$ and average degree at least $d' \geq (d^*)^3/k^2$. As it is shown in [5] that $\min(r_1, r_2, r_3) \leq 2n^{1/3}$, we conclude that $r_4(k, d^*, d_H) \leq d^* 2(n')^{1/3}/d' \leq \frac{2k^2(2d_H)^{1/3}}{(d^*)^2}$.

### 5.3. **Algorithm $A_5$.**

**Lemma 5.7** ([5])**.** *The approximation ratio of Algorithm $A_5$ from [5] is $r_5(k, d^*, d_H) = \Theta(\max(\frac{k^{0.6}d_H^{1.6}}{d^{*2}}, \frac{k^{1/3}d_H^{2/3}}{d^{*2/3}}))$ for the case where $d_H^2 < k$ and $r_5(k, d^*, d_H) = \Theta(\max(\frac{k^{0.4}d_H^2}{d^{*2}}, \frac{d_H^{4/3}}{d^{*2/3}}))$ where $d_H^2 > k > d_H$*

In [5], algorithm $A_5$ was analyzed for a very specific set of parameters $k$, $d^*$, and $d_H$. We now extend the analysis of [5] to obtain the ratio $r_5$ which is a parametric version of the ratio obtained in [5]. In what follows we revisit Procedure 5 (walks of length 5) from [5]. First let us recall a lemma presented and proved in [5]. An $l$-walk between $u$ and $v$ is a path of length $l$ edges between $u$ and $v$.

**Lemma 5.8** ([5])**.** *For a graph with size $k$, average degree $d$ and a number $l$, there must exist at least two vertices $u$ and $v$ for which the number of $l$-walks from $u$ to $v$ is $W_l[u, v] \geq d^l/k$.*

We use the above lemma with $l = 5$. Namely, from the existence of an optimal subgraph $G^*$ of size $k$ and average degree $d^*$ we infer that there must exist at least two vertices $u$ and $v$ in the graph for which $W_5[u, v] \geq (d^*)^5/k$.

We will use this observation in order to calculate the ratio in this case. Let us define $N_1, N_2, N_3, N_4$ to describe the subsets of vertices that are first, second, third, and fourth on the paths of length 5 between $u$ and $v$. As stated in [5], it may be the case that a vertex appears in more than a single set. This may cause some edges in the analysis below to be counted several times - but the multiplicity remains constant, and does not effect the analysis beyond a constant factor. Thus we assume in our analysis that the subsets are disjoint. For completeness, we rewrite the second part of Section 4.2 in [5] with intention to extract the function $r_5$.

*Proof of Lemma 5.7.* Let $\epsilon(d^*, d_H, k)$ be a function to be determined later in the proof. In what follows we will show how to find in $G$ a subgraph $G'$ of size $O(k)$ with average degree $\Omega(\epsilon)$. Using Lemma 5.4 and Theorem 4.2 this implies an approximation ratio for the D$k$S problem of $O(d^*/\epsilon)$. We refer the reader to Remark 4.6, and note that $d_H$ and $k$ do not change during Algorithm 4.3. Our proof is done by a detailed case analysis based on that presented in [5]: Our calculation is seperated into two main cases: $d_H^2 \leq k$ and $d_H^2 \geq k \geq d_H$.

Case 1: $d_H^2 \leq k$. Assume that $cut(N_2, N_3) \geq \epsilon d_H^2$. Here $cut(A, B)$ is the number of edges with one end point in $A$ and another in $B$. In this case, as the size of $N_2$ and $N_3$ is bounded by $d_H{}^2 \leq k$, we take $G'$ to be the graph induced by $N_2 \bigcup N_3$. It holds that $G'$ has average degree $\epsilon/2$. We thus continue under the assumption that $cut(N_2, N_3) < \epsilon d_H{}^2$.

Now assume that there exists a $w \in N_2$ such that $W_3(w, v) > d_H \epsilon$. Observe that all the length 3 walks between $w$ and $v$ must pass through $N_3$ and $N_4$. Consider the graph induced by the neighbors of $w$ in $N_3$, and the set $N_4$. Since $w$ has at most $d_H$ neighbors in $N_3$, this graph contains at most $2d_H$ vertices, and $\Omega(d_H \epsilon)$ edges. Implying an average degree of $\Omega(\epsilon)$. We thus continue under the assumption that for every $w \in N_2, W_3(w, v) \leq d_H \epsilon$ and for every $w \in N_3, W_3(u, w) \leq d_H \epsilon$ (here we use the symmetry of these two assumptions).

We now show, in the setting at hand, that we may also assume that every edge between $N_2$ and $N_3$ lies in at least $\frac{d^{*5}}{2d_H^2 \epsilon k}$ walks from $u$ to $v$ (without loosing much). Indeed, remove any edge between $N_2$ and $N_3$ that lies in less then $\frac{d^{*5}}{2d_H^2 \epsilon k}$ length 5 walks from $u$ to $v$. Since the number of edges between $N_2$ and $N_3$ is less then $d_H^2 \epsilon$, (by our first assumption) we will disconnect at most $(\frac{d^{*5}}{2d_H^2 \epsilon k}) d_H^2 \epsilon = \frac{d^{*5}}{2k}$ walks. So we remain with $\Omega(\frac{d^{*5}}{k})$ walks in $W_5(u, v)$.

Let $e = (w, z)$ be an arbitrary edge between $w \in N_2$ and $z \in N_3$. By our asumptions, $e$ lies in $p \geq \frac{d^{*5}}{2d_H^2 \epsilon k}$ walks from $u$ to $v$. Clearly $p \leq deg(w, N_1) deg(z, N_4)$. Thus, either $deg(w, N_1) \geq \sqrt{\frac{d^{*5}}{2d_H^2 \epsilon k}}$ or $deg(z, N_4) \geq \sqrt{\frac{d^{*5}}{2d_H^2 \epsilon k}}$. If the first one is true, we say $w$ is a 'good' vertex, otherwise $z$ is the 'good' one. Now we activate the following procedure:

1) we choose for every edge $e$, its good vertex (denoted by $w$).

2) we put that vertex in $S_2$ or $S_3$ respectively if it comes from $N_2$ or $N_3$.

3) we remove all the edges that touch $w$ from the graph.

Let $w$ be a good vertex in $N_2$ (a similar analysis holds for good vertices in $N_3$). By our second assumption, $W_3(w, v) \leq d_H \epsilon$. Observe that in step 3 above, we only discard the length 5 walks from $u$ to $v$ that go through $w$. The number of walks between $u$ and $w$ (which equals $degree(w, N_1)$) is bounded above by $d_H$. The number of walks between $u$ and $v$ passing through $w$ is thus bounded by $d_H \epsilon d_H = d_H^2 \epsilon$.

Since we have $\Omega(\frac{d^{*5}}{k})$ walks between $u$ and $v$, and each iteration removes at the most $d_H^2 \epsilon$ of them, we can fix the number of iterations to be $\frac{d^{*5}}{cd_H^2 \epsilon k}$ for a constant $c \geq 1$. Thus the total number of 'good' vertices, found by the algorithm is $\frac{d^{*5}}{cd_H^2 \epsilon k}$.

W.l.o.g assume that $|S_2| \geq |S_3|$. Now consider the subgraph induced by $S_2 \bigcup N_1$. This subgraph contains $O(d_H + \frac{d^{*5}}{cd_H^2 \epsilon k})$ vertices out of which $\frac{d^{*5}}{cd_H^2 \epsilon k}$ vertices have degree at least $\sqrt{\frac{d^{*5}}{2d_H^2 \epsilon k}}$. Thus the average degree of this subgraph is at least

$$\Omega \left( \frac{\frac{d^{*5}}{cd_H^2 \epsilon k} \sqrt{\frac{d^{*5}}{2kd_H^2 \epsilon}}}{d_H + \frac{d^{*5}}{cd_H^2 \epsilon k}} \right)$$

It follows from basic computations that for a constant $c'$, setting $\epsilon$ to be

$$c' \cdot \min(\frac{d^{*3}}{k^{0.6}d_H^{1.6}}, \frac{d^{*5/3}}{k^{1/3}d_H^{2/3}}))$$

the above average degree is at least $\epsilon$. Notice also that since we assumed $k \geq d_H{}^2$ then $|S_2| \leq d_H{}^2 \leq k$ namely, the size of the group we selected is at most $k$.

All in all, in all the sub-cases specified above we obtain a subgraph $G'$ of size at most $k$ with average degree $\Omega(\epsilon)$.

Case 2: $d_H^2 \geq k \geq d_H$. Assume that $cut(N_2, N_3) \geq \frac{d_H{}^4\epsilon}{k}$. We can construct a subset of (expected) size at most $k$ by picking each vertex in $N_2 \bigcup N_3$ with probability $k/2d_H{}^2$. Thus, using Lemma 4.5, we may obtain a subgraph with at most $k$ vertices and $k\epsilon/4$ edges. We thus continue under the assumption that $cut(N_2, N_3) < \frac{d_H{}^2\epsilon}{k}$.

Now assume that there exists $w \in N_2$ with $W_3(w, v) > d_H\epsilon$. Observe that all the length 3 walks between $w$ and $v$ must pass through $N_3$ and $N_4$. Consider the graph induced by the neighbors of $w$ in $N_3$, and the set $N_4$. Since $w$ has at most $d_H$ neighbors in $N_3$, this graph contains at most $2d_H + 1$ vertices, and $\Omega(d_H\epsilon)$ edges. Implying a subgraph of size less than or equal to $k$ with an average degree of $\Omega(\epsilon)$. We thus continue under the assumption that for every $w \in N_2, W_3(w, v) \leq d_H\epsilon$ and for every $w \in N_3, W_3(u, w) \leq d_H\epsilon$ (here we use the symmetry of these two assumptions).

We now show, in the setting at hand, that we may also assume that every edge between $N_2$ and $N_3$ lies in at least $\Omega(\frac{d^{*5}}{d_H^4\epsilon})$ walks from $v$ to $u$.

Indeed, remove any edge between $N_2$ and $N_3$ that lies in less then $\frac{d^{*5}}{2d_H^4\epsilon}$ length 5 walks from $v$ to $u$. Since the number of edges between $N_2$ and $N_3$ is less then $\frac{d_H^4\epsilon}{k}$, (by our first assumption) we will disconnect at most $(\frac{d^{*5}}{2d_H^4\epsilon})\frac{d_H^4\epsilon}{k} = \frac{d^{*5}}{2k}$ walks in $W_5(u, v)$, so we remain with $\Omega(\frac{d^{*5}}{k})$ walks (recall that $W_5(u, v) \geq \frac{d^{*5}}{k}$).

Let $e = (w, z)$ be an arbitrary edge between $w \in N_2$ and $z \in N_3$. By our assumptions, $e$ lies in $p \geq \frac{d^{*5}}{2d_H^4\epsilon}$ walks from $u$ to $v$. Clearly $p \leq deg(w, N_1)deg(z, N_4)$. Thus, either $deg(w, N_1) \geq \sqrt{\frac{d^{*5}}{2d_H^4\epsilon}}$ or $deg(z, N_4) \geq \sqrt{\frac{d^{*5}}{2d_H^4\epsilon}}$. If the first one is true, we say $w$ is a 'good' vertex, otherwise $z$ is the 'good' one. Now we use the same analysis and procedure as in the corresponding case 1. Namely, we construct the sets $S_2$ and $S_3$. As before, for every $w \in N_2$, by our assumption above, $W_3(w, v) \leq d_H\epsilon$. Moreover, in our procedure we only discard the length 5 walks from $v$ to $u$ that go through $w$. Since $W_3(w, v) \leq d_H\epsilon$ and since the number of walks between $u$ and $w$ (which equals $deg(w, N_1)$) is bounded above by $d_H$, it implies that the number of walks between $u$ and $v$ passing through $w$, is bounded by $d_H \epsilon d_H = d_H^2 \epsilon$.

Since we have $\Omega(d^{*5}/k)$ walks between $v$ and $u$, and each iteration removes only $d_H^2 \epsilon$ of them, the number of iterations can be at least $\Theta(\frac{d^{*5}}{d_H^2 \epsilon k})$. However, if this number is greater than $k$, we will stop after $k$ iterations only. We proceed under the assumption that $\Theta(\frac{d^{*5}}{d_H^2 \epsilon k}) \leq k$, i.e. the total number of 'good' vertices, found by the procedure is $\Theta(\frac{d^{*5}}{d_H^2 \epsilon k})$. We deal with the case that $\Theta(\frac{d^{*5}}{d_H^2 \epsilon k}) \geq k$ at the end of the proof.

W.l.o.g assume that $|S_2| \geq |S_3|$. Now consider the subgraph induces by $S_2 \bigcup N_1$. It contains $O(d_H + \frac{d^{*5}}{\epsilon d_H^2 k})$ vertices out of which $\Theta(\frac{d^{*5}}{\epsilon d_H^2 k})$ vertices have degree at least $\sqrt{\frac{d^{*5}}{2\epsilon d_H^4}}$. Thus the average degree of this subgraph is at least

$$\Omega\left(\frac{\frac{d^{*5}}{k d_H^2 \epsilon} * \sqrt{\frac{(d^*)^5}{d_H^4 \epsilon}}}{d_H + \frac{d^{*5}}{\epsilon d_H^2 k}}\right)$$

It follows from basic computations that for a constant $c'$, setting $\epsilon$ to be

$$\epsilon = \Omega(\min(\frac{d^{*3}}{k^{0.4} d_H^2}, \frac{d^{*5/3}}{d_H^{4/3}}))$$

the above average degree is at least $\epsilon$.

In case $\Theta(\frac{d^{*5}}{d_H^2 \epsilon k}) \geq k$ we have that $S_2$ is of size $O(k)$ (as we stopped the iterative process early). Thus, as $k \geq d_H$, in this case we obtain an average degree $\epsilon'$ of

$$\epsilon' = \Omega\left(\frac{k\sqrt{\frac{(d^*)^5}{d_H^4 \epsilon}}}{d_H + k}\right) \geq \Omega\left(\frac{k\sqrt{\frac{(d^*)^5}{d_H^4 \epsilon'}}}{2k}\right) = \Omega\left(\frac{(d^*)^{5/3}}{d_H^{4/3}}\right) \geq \epsilon$$

So limiting our subgraph to $k$ does not change the overall ratio.

So, all in all (in case 1 and 2 analyzed above), we have found a subgraph with size smaller or equal to $k$ with average degree $\Omega(\min(\frac{d^{*3}}{k^{0.6} d_H^6}, \frac{d^{*5/3}}{k^{1/3} d_H^{2/3}}))$ for the case where $d_H^2 \leq k$ and $\Omega(\min(\frac{d^{*3}}{k^{0.4} d_H^2}, \frac{d^{*5/3}}{d_H^{4/3}}))$ when $d_H^2 > k > d_H$. The size being smaller or equal to $k$ implies a solution for the Dam$k$S problem. But as we have shown in Section 4, there is a reduction between the the Dam$k$S and the D$k$S problem, with only a constant lost of approximation. Thus we use this calculated average degree to get the ratio $r_5$ stated in Lemma 5.7. $\qquad \square$

### 5.4. Proof of Theorem 5.1: Combining algorithms $A_1, \ldots, A_5$.

Using Lemmas 5.2-5.7 we may now compute a lower bound to the approximation ratio of the combined algorithm of [5]: $\max_{k,d^*,d_H} \min_{i=1,\ldots,5} r_i(k, d^*, d_H)$. We do this numerically going over many triplets $(k, d^*, d_H)$ and computing $\min_{i=1,\ldots,5} r_i(k, d^*, d_H)$ for each such triplet. This suffices to yield the lower bound in Theorem 5.1 for the ratio of [5]. Our C-program which preforms the computation above is given in Section 7.1 and runs in precision 0.00001. In Section 6 we show that we can improve upon this ratio.

## 6. Improving on the approximating of the D$k$S

As we have shown in the previous section, prior to our work, the best approximation ratio for the Dense-$k$-Subgraph problem stands on $n^\delta$ for a constant $\delta \geq 0.32258$. Improving upon this approximation ratio is a long standing open problem.

In [5] a total of 5 algorithms are presented. The first three algorithms of [5] tend to work 'well' in several cases. For example, when we work on very dense graphs, where $d_H$ - the average degree of the $k/2$ vertices with the highest degree in $G$, is high relatively to $n$. Other cases include the case in which $k$ itself is very large relatively to $n$ or when $d^*$ the average degree of the optimal subgraph is high relatively to $\max(k, d_H)$ and/or $n$. When each of these parameters are very small - less then $n^{1/3}$, the algorithm gets good approximation for trivial reasons. Two worst cases for these algorithms where isolated in [5], in which the analysis gave exactly an $O(n^{1/3})$ approximation:

$$(6.1) \qquad d^*(G, k) = \theta(n^{1/3}), \ k = \theta(n^{1/3}), \ d_H = \theta(n^{2/3}).$$

$$(6.2) \qquad d^*(G, k) = \theta(n^{1/3}), \ k = \theta(n^{2/3}), \ d_H = \theta(n^{1/3}).$$

where $d_H$ is the average degree of the $k/2$ vertices with the highest degree in $G$. To improve upon this ratio, [5] suggest two additional algorithms, $A_4$ and $A_5$, one for each isolated region presented above. The result is the approximation ratio stated in Theorem 5.1. In what follows we present yet an additional algorithm $A_6$ tailored to improve the ratio on the configurations of $(k, d^*, d_H)$ in which the algorithms of [5] work poorly. More specifically, our algorithm is designed to address the region in Equation 6.2 above.

Analyzing the ratio of our additional algorithm $A_6$, we are able to prove the following theorem:

**Theorem 6.1.** *The DkS problem admits an approximation ratio of at most $n^{0.3159}$.*

The proof of Theorem 6.1 includes four steps. First, recall by Lemma 5.4 that one may assume w.l.o.g. that the input graph has maximum degree $d_H$. Then we turn to present our algorithm $A_6$. We start by presenting in Section 6.1 an algorithm for the Dam$k$S problem, that does not return a dense subgraph of size $k$ but rather of size *at most* $k$. Our algorithm is based on a linear programming relaxation and assumes that the given input graph has bounded degree (which as discussed is w.l.o.g.). To obtain our algorithm $A_6$ for D$k$S, we use our analysis presented in Section 4 in which we show that any algorithm for Dam$k$S implies an algorithm for D$k$S with roughly the same approximation ratio. Finally, in Section 7 we analyze the approximation ratio of algorithms $A_1, A_2, A_3, A_4$ combined with our $A_6$.

$$\max_{k, d^*, d_H} \min_{i=1,\ldots,4,6} r_i(k, d^*, d_H).$$

Here, as our calculations are numerical, we need to bound any error term that may arise due to the fact that we are not going over all possible triplets $(k, d^*, d_H)$ but rather only checking a *grid* of triplets at limited precision. We bound the error term by 0.0000433 in Section 7. The results of Theorem 6.1 takes this error term into account. Our numerical calculations (i.e. our C program) appear in the Section 7.1.

6.1. **Our algorithm for Dam$k$S.** Our algorithm is based on a linear program (LP). We use the fractional LP solution in order to find a Dam$k$S solution with an average degree of $\Omega((\frac{d^{*7}}{d_H^4 k})^{\frac{1}{3}})$.

6.1.1. *The DamkS LP.* Let $G(V, E)$ be a given graph and let $V = 1, \ldots, n$. Let $i_0$ be some vertex in $V$, and let $\gamma$ be a parameter. Our linear problem for Dam$k$S follows:

$$(6.3) \qquad\qquad\qquad min \sum y_i$$

$$(6.4) \qquad\qquad\qquad st : y_{i_0} = 1$$

$$(6.5) \qquad\qquad \forall i \in V \quad y_i \leq \sum_{j:(i,j)\in E} x_{ij}/\gamma$$

$$(6.6) \qquad\qquad\qquad x_{ij} \leq y_i$$

$$(6.7) \qquad\qquad\qquad x_{ij} \leq y_j$$

$$(6.8) \qquad\qquad \forall i \quad y_i \in [0,1]$$

In what follows, let $d^*$ be the average degree of the optimal solution to the Dam$k$S problem.

**Lemma 6.2.** *There exists a vertex $i_0$ such that solving the DamkS LP with $\gamma = d^*/2$ yields an optimal solution of value at most $k$.*

*Proof.* Consider the optimal solution $U \subset V$ to the Dam$k$S problem on $G$. Let $d^*$ be the average degree of $U$. In Lemma 2 of [1] it is shown that $U$ must include a subgraph $U'$ of minimal induced degree at least $d^*/2$. Let $i_0$ be a vertex in $U'$. Setting $y_i = 1$ for each vertex in $U'$ and 0 otherwise, and setting $x_{ij} = 1$ for each edge $(i, j)$ in $U$ and 0 otherwise, yields a valid solution to our LP of value $|U'| \leq k$. This implies our assertion. □

In what follows we assume that we have guessed the correct value for $i_0$ and $\gamma$ as stated in the Lemma above. To assure this in practice we will run our algorithm (yet to be defined) for all possible values of $i_0$ (there are $n$ such values) and for all values of $\gamma$ in the set $\{1, 2, 4, 8, \ldots, n\}$, taking our algorithm to be the *best* out of these $n \log n$ executions. As we are approximating $\gamma$ within a factor of 2, we obtain a slight loss in our approximation ratio which (as we will see) can be bounded by a constant value of 4 (and is thus insignificant to our results). We also assume that $\gamma$ is at least $n^{0.01}$, otherwise $A_1$ yields an excellent approximation ratio. Finally, in what follows we refer to the variable $y_{i_0}$ as $y_0$.

The Dam$k$S LP gives weights $y_i$'s to the vertices. Next we present an algorithm that uses these weights in order to extract a subgraph with at most $k$ vertices.

6.1.2. *The algorithm.* We start with some definitions. Let $N_0$ be the set consisting of a single vertex corresponding to $y_0$ in our LP. Refer to this vertex as $v_0$. Let $N_1$, $N_2$ and $N_3$ be the groups of vertices that are of distance one, two, and three from $v_0$ respectively. Here the distance between two vertices is the length of the shortest path connecting them. Notice that for every $i \neq j$: $N_i \cap N_j = \phi$.

Let $d_H$ be the maximum degree in the input graph. In our algorithm we consider two candidate subgraphs.

The subgraph $S_1$ is constructed by randomly picking vertices $i$ from $N_0 \cup N_1 \cup N_2$. Each vertex $v_i$ is picked independently with its corresponding probability $y_i$. The subgraph $S_2$ is constructed by randomly picking vertices $v_i$ from $N_1 \cup N_2 \cup N_3$, again each vertex is picked with its corresponding probability $y_i$. For our analysis, we assume that the random choices for $S_1$ and $S_2$ are *fresh* (i.e. independent). We choose the densest between the two subgraphs.

First we prove a technical lemma to be used later in our proof.

**Lemma 6.3.** $\sum_{1,\ldots,n} y_i^2 \geq \frac{(\sum_{1,\ldots,n} y_i)^2}{n}$

*Proof.* By the Cauchy–Schwartz inequality we get

$$\sum_{1,\ldots,n} y_i = \sum_{1,\ldots,n} 1 y_i \leq \sqrt{\sum_{1,\ldots,n} 1^2} \sqrt{\sum_{1,\ldots,n} y_i{}^2} = \sqrt{n \sum_{1,\ldots,n} y_i{}^2}$$

. If we square both sides and revise the equation, we get the lemma result. $\square$

**Lemma 6.4.** *Algorithm $A_6$ yields an approximation ratio of $r_6 = \Theta\left(\frac{d^4 k}{\gamma^4}\right)^{1/3} = \Theta\left(\frac{d^4 k}{d^{*4}}\right)^{1/3}$.*

*Proof.* Let us call $Q_0, Q_1, Q_2$ and $Q_3$ the sums of the $y_i$ values in $N_0, N_1, N_2$ and $N_3$ respectively. By the LP feasibility of the $y_i$'s and due to constraints 6.4, 6.5, 6.6, and 6.7 we conclude $Q_1 \geq \gamma$. We calculate the average degree of $S_1$.

We start by bounding the expected number of edges in $S_1$ from below. To this end, we calculate the expected sum of degrees of vertices in $S_1 \cap N_1$. Notice that each edge $(i, j)$ adjacent to $N_1$ will appear in the subgraph induced by $S_1$ with probability $y_i y_j$. Thus:

$$(6.9) \qquad \sum_{i \in N_1}\left(y_i \sum_{ij \in E} y_j\right) \geq \sum_{i \in N_1} y_i{}^2 \gamma \geq \frac{Q_1{}^2}{d_H} \gamma$$

Here we used Lemma 6.3 in the last inequality (and the fact that $|N_1| \leq d_H$). Thus we have that the expected number of edges in $S_1$ is at least $\frac{Q_1{}^2}{2 d_H} \gamma$

We now turn to study the expected size of $S_1$. This expectation is exactly $Q_0 + Q_1 + Q_2$ (which is at most $k$). Thus if both the number of edges in $S_1$ and its size behave as expected, we will obtain a subgraph $S_1$ of size $\leq k$ with average degree $d_1 = \frac{Q_1{}^2 \gamma / d_H}{Q_0 + Q_1 + Q_2}$. Now if $Q_2 \leq 2 Q_1$ then we obtain $d_1 = \Theta\left(\frac{Q_1 \gamma}{d_H}\right) \geq \Theta\left(\frac{\gamma^2}{d_H}\right)$ which is an excellent ratio. Otherwise, $Q_2 > 2 Q_1$, and hence:

$$(6.10) \qquad d_1 = \frac{Q_1{}^2 \gamma / d_H}{Q_0 + Q_1 + Q_2} = \Theta\left(\frac{Q_1{}^2 \gamma}{d_H Q_2}\right) \geq \Theta\left(\frac{\gamma^3}{d_H Q_2}\right)$$

It is left to show that with some polynomial probability it is indeed the case that both the size and the number of edges in $S_1$ behave as expected. The number of edges in $S_1$, is bounded by $n^2$ and thus using Markov's inequality it holds that this number will be at least half its expectation with probability at least $\frac{1}{2n^2}$. Regarding the number of vertices in $S_1$, as these are chosen independently, using the Chernoff bound it holds that

$$\Pr[|S_1| \leq 2(Q_0 + Q_1 + Q_2)] \geq 1 - e^{-\frac{Q_0 + Q_1 + Q_2}{3}}$$

As $Q_0 + Q_1 + Q_2 \geq Q_1 \geq \gamma \geq n^{0.01}$ (the latter by our assumption discussed at the beginning of this section), it holds that (for sufficiently large $n$) both the size of $S_1$ and the number of edges it induces are within a factor of 2 from their expectation with probability at least $\frac{1}{2n^2} - e^{\frac{-n^{0.01}}{3}} \geq \frac{1}{4n^2}$.

We now address $S_2$. As before the expected sum of degrees in the subgraph induced by $S_2$ is at least $\sum_{i \in N_2}(y_i \sum_{ij \in E} y_j) \geq \sum_{i \in N_2}(y_i{}^2 \gamma) \geq \frac{Q_2{}^2}{d_H{}^2}\gamma$. Again we used Lemma 6.3 for the last inequality (here, $|N_2| \leq d_H^2$). We divide the expected sum of degrees with the expected size of $S_2$ to get

$$d_2 = \frac{\gamma Q_2{}^2 / d_H{}^2}{Q_1 + Q_2 + Q_3} > \frac{\gamma Q_2{}^2}{d_H^2 k}.$$

In the last inequality we used the fact that $k \geq LP \geq Q_0 + Q_1 + Q_2 + Q_3 > Q_1 + Q_2 + Q_3$ (which also implies that in expectation $|S_2| \leq k$).

Again we show that the size and the number of edges in $S_2$ behave as expected with some non-negligible probability. The number of edges in $S_2$, is bounded by $n^2$ and thus using Markov's inequality it holds that this number will be at least half its expectation with probability $\frac{1}{2n^2}$. Regarding the number of vertices in $S_2$, as these are chosen independently, using the Chernoff bound it holds that

$$\Pr[|S_2| \leq 2(Q_1 + Q_2 + Q_3)] \geq 1 - e^{-\frac{Q_1 + Q_2 + Q_3}{3}}$$

As $Q_1 + Q_2 + Q_3 \geq Q_1 \geq \gamma \geq n^{0.01}$ , it holds that (for sufficiently large $n$) both the size of $S_2$ and the number of edges it induces are within a factor of 2 from their expectation with probability $\frac{1}{2n^2} - e^{\frac{-n^{0.01}}{3}} \geq \frac{1}{4n^2}$.

Thus, all is all, with probability at least $\frac{1}{16n^4}$ both the average degree of $S_1$ and $S_2$ are at least $\Omega(d_1)$ and $\Omega(d_2)$ respectively, and both $S_1$ and $S_2$ are at most of size $2k$. Reducing the size of $S_1$ and $S_2$ (if needed) by Lemma 4.5 will not change the value of $d_1$ or $d_2$ significantly. Notice that when $Q_2$ increases then so does $d_2$, while when $Q_2$ decreases then $d_1$ is increased. It now follows (by picking the worst value of $(d_H k \gamma^2)^{\frac{1}{3}}$ for $Q_2$) that we are guaranteed an average degree of $\Theta(\frac{\gamma^7}{d_H^4 k})^{\frac{1}{3}}$ with the above probability. Repeating the above rounding procedure a polynomial number of times (say $n^5$) and taking the densest obtained subgraph we conclude that the above average degree is obtained with arbitrarily high (constant) probability. This suffices to prove our lemma. $\qquad \square$

We have integrated the results of this section in our C program that calculates the final ratio of combining algorithms $A_1, A_2, A_3, A_4$ and our $A_6$. The resulting ratio is, $r = 0.3159$. Our detailed calculations appears at Section 7.

## 7. COMPUTING THE FINAL RATIO USING A C PROGRAM

Our C program runs on a grid of values for $d^*$,$d_H$ and $k$ and computes the resulting approximation ratio (as a function of $n$) for every point in this three dimensional grid. To be more specific, the set of values for $d^*$ (and also $d_H$ and $k$) has an exponentially growing nature and consists of the set $\{n^i \mid i = 0, \Delta, 2\Delta, \ldots, 1\}$ of size $1/\Delta$ for a precision parameter $\Delta$. The worst case setting of values for $d^*$,$d_H$ and $k$ can (and probably is) located between the grid points and not necessarily on one of them. To compensate for this fact, we analyze the maximum loss in our ratio that may occur. Technically, as our C program computes the exponent of $n$ in the resulting ratio, our loss can be computed using the linear nature of our equations defining the (logarithm of the) ratio $r_i$ of each of the algorithms $A_i$. For example, consider $r_2 = \frac{2nd^*}{kd_H}$. Neglecting the factor of 2 and setting $d^* = n^g$, $d_H = n^d$, and $k = n^K$, we have that $r_2 = n^{g-K-d+1}$. Thus, the total loss in considering our grid of precision $\Delta$ in this case will be $n^{3\Delta}$ (a single factor of $\Delta$ for each one of $g$, $k$, and $d$). Let us compute our total error $\delta_{err}$ corresponding to the value of $\Delta$ we are using.

The approximation ratio of Algorithm $A_1$ is $r_1(k, d^*, d_H) = d^*$. In the C-program this translates into $n^g$, so $\delta_1 = 1 \cdot \Delta$. The approximation ratio of Algorithm $A_2$ is $r_2(k, d^*, d_H) = \frac{2nd^*}{kd_H}$. In the C-program this translates into $n^{g-K-d+1}$, so $\delta_2 \leq 3\Delta$. The approximation ratio of Algorithm $A_3$ is $r_3(k, d^*, d_H) = 2max(k, d_H)/d^*$. In the C-program this translates into $n^{max(K,d)-g}$, so $\delta_3 \leq 2\Delta$. The approximation ratio of Algorithm $A_4$ is $r_4(k, d^*, d_H) = \frac{2k^2 d_H^{1/3}}{(d^*)^2}$. In the C-program this translates into $n^{2K+\frac{1}{3}d-2g}$, so $\delta_4 \leq 4\frac{1}{3}\Delta$. When using Algorithm $A_6$ in our program, algorithm $A_5$ is not needed (as it does not contribute to improving the approximation ratio), hence we don't have to consider $\delta_5$. The approximation ratio of Algorithm $A_6$ is $r_6(k, d^*, d_H) = (\frac{d_H^4 k}{(d^*)^4})^{\frac{1}{3}}$. In the C program it translates into $n^{\frac{4}{3}d+\frac{1}{3}K-\frac{4}{3}g}$ so $\delta_6 \leq 3\Delta$. Our total error $\delta_{error}$ is thus bounded by $max(\delta_1, \delta_2, \delta_3, \delta_4, \delta_6) = 4\frac{1}{3}\Delta$

Running our C program while considering only the algorithms $A_1, \ldots, A_5$ of [5] with $\Delta = 0.00001$, we receive the ratio $r_{previous} = 0.32258$. This suffices to prove Theorem 5.1 as all we seek is a lower bound on the ratio derived from the stated values of $r_1, \ldots, r_5$.

Running our C program with the algorithms of [5] and our additional algorithm $A_6$ with $\Delta = 0.00001$, we receive the ratio $0.315787$. Adding the error $\delta_{err} = 0.0000433$ results in the ratio of $r_{new} \leq 0.3159$ stated in Theorem 6.1.

### 7.1. **Our C program for computing the approximation ratio.** Our program appears in Figure 1.

## References

[1] R. Anderson and K. Chellapilla. Finding Dense Subgraphs with Size Bounds. *In proceedings of the 6th Workshop on Algorithms and Models for the Web Graph (WAW2009), pages 25-37.*

[2] Y. Asahiro, R. Hassin and K. Iwama. Complexity of finding dense subgraphs, *Discrete Appl. Math. 121(1-3), pp. 1526,* 2002.

[3] A. Bhaskara, M. Charikar, E. Chlamtac, U. Feige and A. Vijayaraghavan. Detecting High Log-Densities an $O(n^{1/4})$ Approximation for Densest k-Subgraph. *CoRR*, abs/1001.2891: 2010.

[4] M. Charikar. Greedy approximation for finding dense components in a graph. *In proceedings of APPROX*, pages 139–152, 2000.

[5] U. Feige, G. Kortsarz, and D. Peleg. The dense k-subgraph problem. *Algorithmica*, 29(3):410–421.

[6] U. Feige and M. Langberg. Approximation algorithms for maximazation problems arising in graph partitioning. *Journal of Algorithms 41, pp. 174–211,* 2001.

[7] U. Feige and M. Seltser. On the densest k-subgraph problem. *Technical report, Department of Applied Mathematics and Computer Science, The Weizmann Institute, Rehobot,* 1997.

[8] A. Goldberg. Finding a maximum density subgraph. *Technical Report UCB/CSB 84/171, Department of Electrical Engineering and Computer Science, University of California, Berkeley, CA,* 1984.

[9] S. Khot. Ruling out PTAS for graph min-bisection, dense k-subgraph, and bipartite clique. *SIAM Journal on Computing, 36(4), pp. 1025-1071,* 2006.

[10] S. Khuller and B. Saha: On Finding Dense Subgraphs. *In proceedings of ICALP*, 597-608: 2009.

[11] A. Schrijver. A Combinatorial Algorithm Minimizing Submodular Functions in Strongly Polynomial Time. *J. Combinatorial Theory, vol. B 80, pp. 346-355,* 2000.

```
main()
{
    double g = 0; // d* = n^g
    double k = 0; // k = n^K
    double d = 0; // d_H = n^d
    double r = 0; // temp ratio
    double rr= 0; // final ratio
    double p = 0.00001; // step
    for (g = 0.0 ; g <= 1 ; g += p ) {
        for ( d = g ; d <= 1 ; d += p ) {
            for ( K = g ; K <= 1 ; K +=p ) {
            // due to algorithm A_1
            r = g - 0;
            // due to A_2
            r = min( r , g-K-d+1 );
            // due to A_3
            r = min( r , g-2*g+max(K,d));
            // due to A_4
            r = min( r , g-3*g+2*K+d/3.0);
            // due to A_5
            if (2*d <= K)
                r = min(r,g - min((3*g-1.6*d-0.6*K),(5.0*g-K-2.0*d)/3.0 ));
            else if (( K < 2*d) && (K > d))
                r = min(r,g - min(3*g-2*d-0.4*K,(5.0*g-4.0*d)/3.0));
            // due to A_6 our LP algorithm
            r = min(r,g-(7.0*g - 4.0*d - K)/3.0);
            if (rr < r) {
                rr=r;
                printf("d=%f K=%f g=%f rr=%f n",d,K,g,rr);
            }
        }
    }
}
```

FIGURE 1. Our C-program that computes the total ratio of all the sub algorithms combined.